\documentclass{eas}
\usepackage{graphicx}
\usepackage{amssymb,amsmath}
\def\arcsec{\hbox{$^{\prime\prime}$}}
\begin{document}

\title{Star formation in infrared dark clouds: Self-gravity and dynamics} 
\author{Nicolas Peretto}\address{School of Physics \& Astronomy, Cardiff University, Cardiff, UK}
\author{Mathilde Gaudel$^{\,1}$}
\author{Fabien Louvet}\address{Departemento de Astronom\'ia, Universidade de Chile, Santiago, Chile}
\author{Gary A. Fuller}\address{Jodrell Bank Centre for Astrophysics, University of Manchester, Manchester, UK}
\author{Alessio Traficante$^{\,3}$}
\author{Ana Duarte-Cabral}\address{School of Physics \& Astronomy, University of Exeter, Exeter, UK}
\begin{abstract}
The role played by gravity in the transfer of interstellar matter from molecular cloud scales to protostellar scales is still highly debated. Only detailed studies on the kinematics of large samples  of star-forming clouds will settle the issue. We present new IRAM 30m observations of a sample of 27 infrared dark clouds covering a large range of sizes, masses, and aspect ratios. Preliminary results suggest that gravity is regulating the dynamical evolution of these clouds on a couple of parsec scales.

\end{abstract}
\maketitle
\section{Introduction}

Since the seventies (Zuckerman \& Evans 1974) it has been recognised that the observed star formation rate in the Milky Way is only a few percents of what we would expect if all molecular clouds were to be in spherical free-fall collapse. This led to the development of a number of theories in which clouds are quasi-statically evolving over several tens of free-fall times, magnetic and/or turbulent pressure preventing them from collapsing on all scales down to core sizes (e.g. Krumholz \& Tan 2007). This picture, though, is not supported by recent results suggesting that the rapid global collapse of massive-star forming clouds is responsible for the formation of very massive cores at the centre of their gravitational potential well (e.g. Peretto et al. 2013). Systematic studies on cloud dynamics are crucially needed to better understand the dynamical evolution of star-forming clouds and how it relates to the star formation rate in the Galaxy. In this context, we have studied a sample of infrared dark clouds, in search  for dynamical patterns that give some insight into  the modes of gravitational collapse at cloud scales.

\section{Source sample and data}

We present the analysis of a sample of 27 infrared dark clouds (IRDCs) taken from the Peretto \& Fuller (2009) IRDC catalogue. They have been selected to lie within a narrow range of kinematical distances, i.e. 3~kpc~$<d<$~5~kpc. We used HIGAL data (Molinari et al. 2010) to compute H$_2$ column density maps towards each of the 27 IRDCs. This was done by performing a 4-point pixel-by-pixel fit of the spectral energy distribution  at 160/250/350/500~$\mu m$. The angular resolution of these column density maps is 36\arcsec. We used these maps along with the clouds' kinematic distances to estimate the masses, sizes, and aspect ratios of all 27 IRDCs (see Fig.~1).
We also mapped each cloud with the IRAM 30m in the N$_2$H$^+$(1-0) and HCO$^+$(1-0) lines. The angular resolution of these observations is 27\arcsec, while the velocity resolution is 0.16~km/s. The rms noise in the IRAM data is $\sim 0.15$~K.  We analysed these IRAM data, in combination with the Herschel data, to investigate the dynamical state of these clouds.

 \begin{figure}
 \hspace{0.5cm}
   \includegraphics[width=10cm]{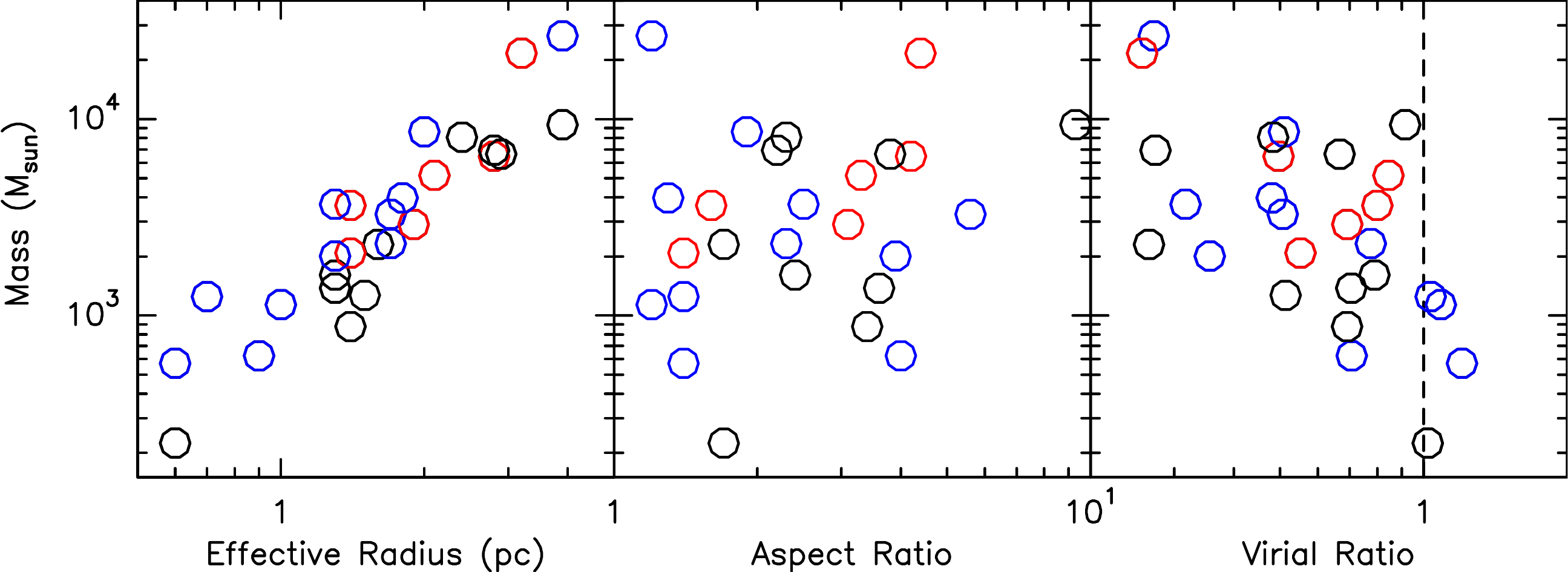}
         \caption{Masses, radii, aspect ratios, and virial ratios for the 27 IRDCs of our sample. Blue, red and black symbols correspond to clouds dominated by HCO$^+$(1-0) blue-shifted, red-shifted, and symmetric line shapes, respectively.}
         \label{spectra}
   \end{figure}

\section{Virial ratios, infall signatures, and free-fall velocities }


We split our clouds into three categories based on the shape of the HCO$^+$(1-0) line emission: i. self-absorbed blue shifted emission; ii. red-shifted emission; iii. symmetric emission. The reason for doing so is that self-absorbed blue shifted emission lines are usually interpreted as signatures of infall, even though the interpretation can be trickier in  clouds with complex morphologies and velocity fields (Smith et al. 2013). Overall, 41\% of our sample is dominated by infall signatures, 22\% by expansion signatures, and 37\% with no specific signatures.

 \begin{figure}
 \hspace{1.4cm}
   \includegraphics[width=4.cm]{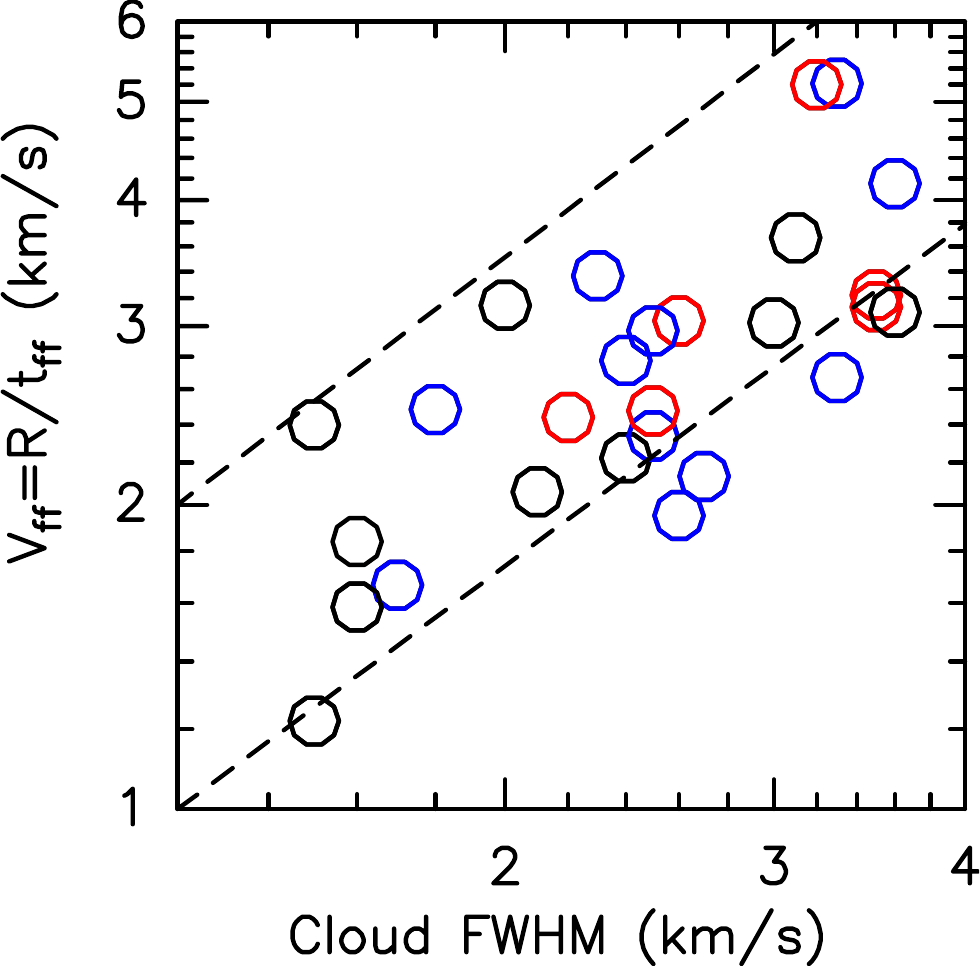}
   \hspace{0.3cm}
           \includegraphics[width=4.1cm]{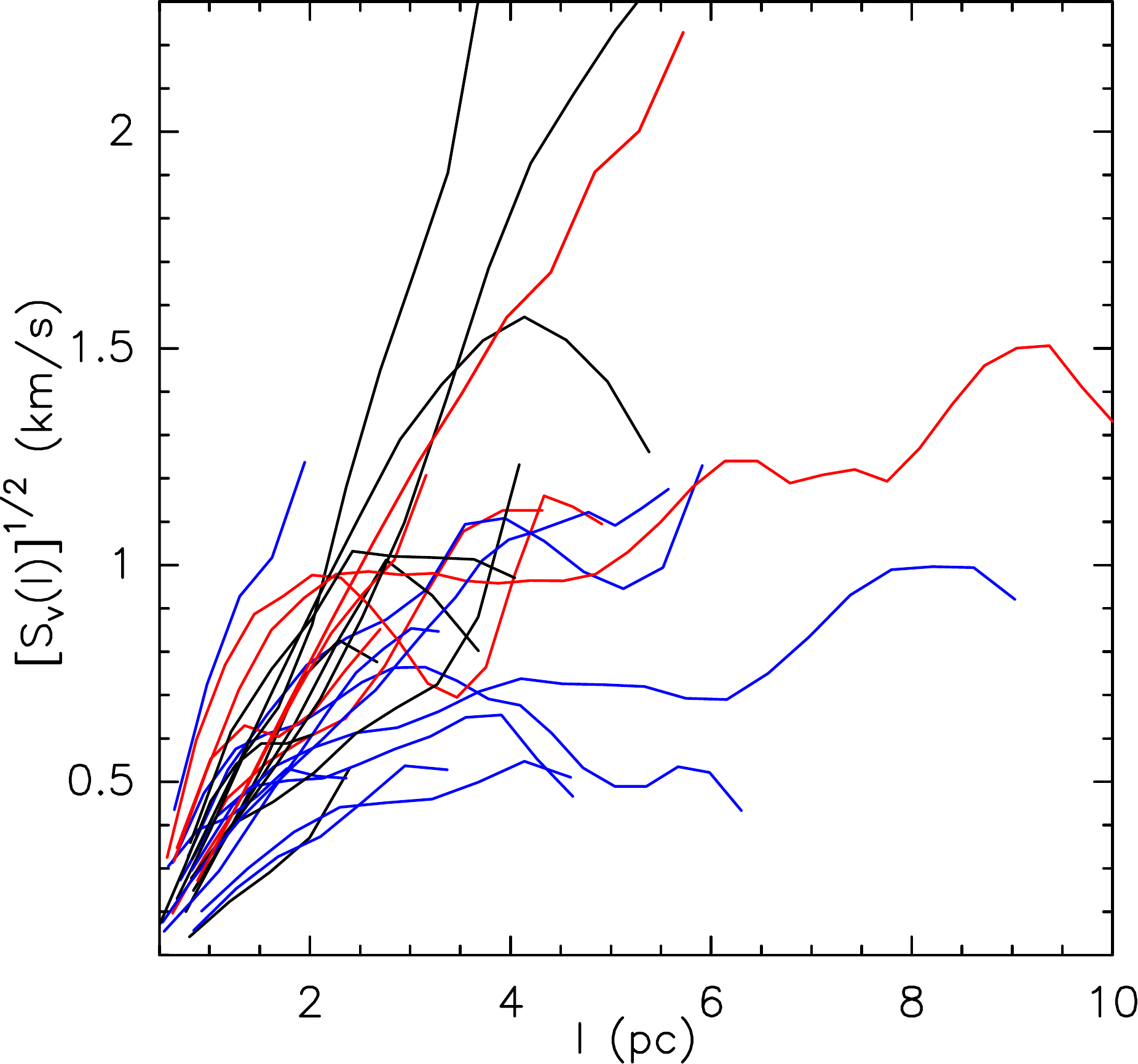}
         \caption{(left): N$_2$H$^+$(1-0) FWHM average over the clouds vs free-fall velocities. The colour coding of the symbols is the same as in Fig.~1. (right): Square root of velocity structure functions for all clouds.}
         \label{spectra}
   \end{figure}

We then used the optically thin N$_2$H$^+$(1-0) line to estimate the velocity dispersion of the IRDCs, averaged over the extent of each cloud, and compute their virial ratios $\alpha_{vir}$ following $\alpha_{vir}=\frac{4R\sigma^2}{GM}$ where $R$ is the radius of the cloud, $\sigma$ is the 1D velocity dispersion, and $M$ is its mass. Here we have assumed a cloud density profile $\rho\propto r^{-1.5}$, consistent with the Herschel observations. Figure 1 shows that the distribution of $\alpha_{vir}$ for our sample is consistent with $\alpha_{vir}\lesssim 1$, implying that all clouds are self-gravitating.  We do not see any clear emerging trend between cloud properties (mass, size, aspect and virial ratios) and HCO$^+$(1-0) infall signatures. This could indicate either that HCO$^+$(1-0) line profiles are not reliable indicators of infall and/or that virial ratios as we estimated them do not tell us much about the dynamical state of clouds.

If low $\alpha_{vir}$ clouds are evolving quasi-statically, then strong magnetic fields must provide the necessary support to prevent their collapse. The field strength which is required is of a few tenths of mG on scales of a couple of parsecs. This is marginally compatible with magnetic field strengths reported in the literature (Crutcher 2012). If the IRDCs of our sample are all collapsing, then we might expect a relationship between the free-fall velocity of the clouds and its observed velocity dispersion. The free-fall velocity $v_{\rm{ff}}$ is computed via $v_{\rm{ff}}=\frac{R}{t_{\rm{ff}}}=\left(\frac{8GM}{\pi^2R}\right)^{1/2}$. Figure~2 (left) displays $v_{\rm{ff}}$ versus the full-width at half-maximum ($=2.35\sigma$) of the N$_2$H$^+$(1-0) line  for all IRDCs. We see that there is a linear correlation between these two quantities pointing towards the possibility that the gas velocity dispersion is driven by the gravitational collapse  itself (see also Traficante et al. in this chapter). However, note that a one-to-one correlation between $v_{\rm{ff}}$ and $\sigma$ means a constant $\alpha_{vir}$ ratio. Therefore, the factor $\sim 2$ dispersion observed in Fig.~2 (left) is mirrored by a factor of $\sim 5$  in dispersion in Fig.~1 (right) since $\alpha_{vir}\propto\sigma^2$.  In the collapse scenario, low $\alpha_{vir}$ clouds could be in an early evolution of collapse.

\begin{figure}
 \hspace{1cm}
   \includegraphics[width=9.cm]{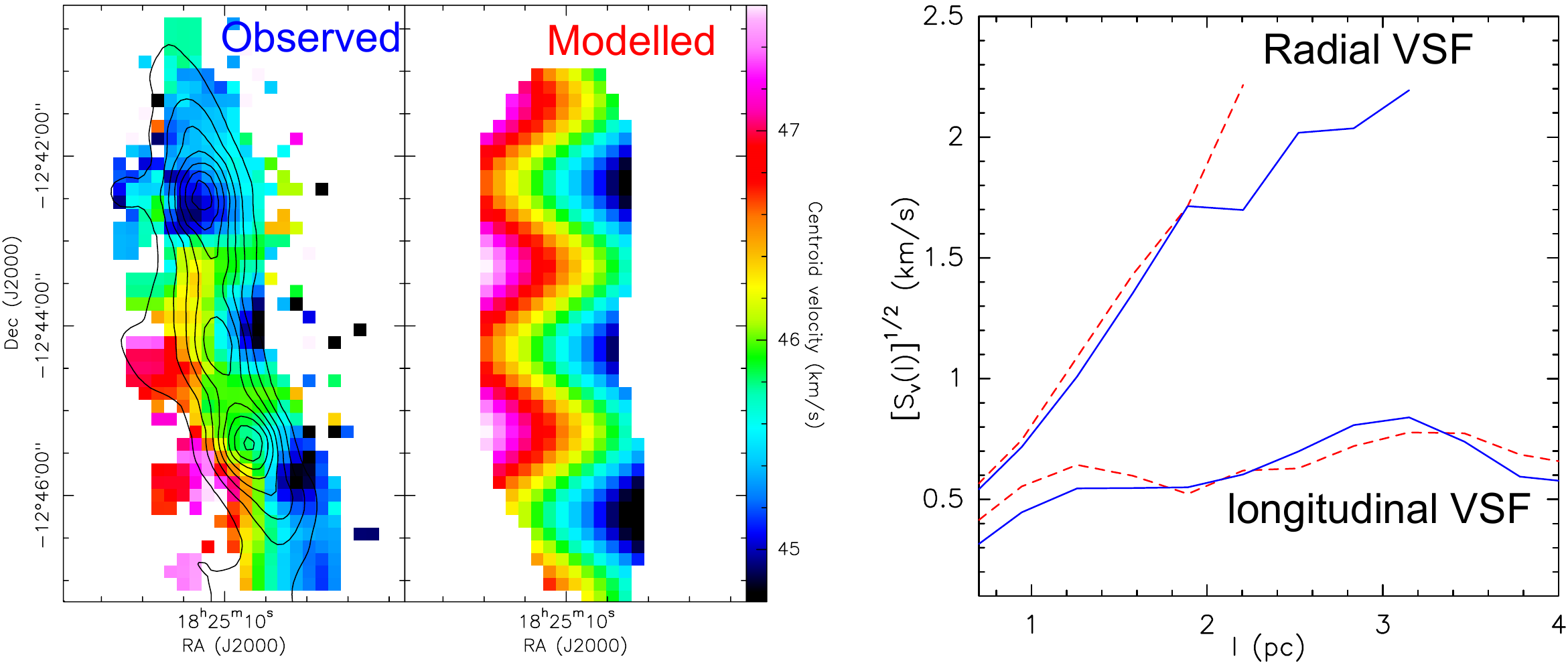}
           \caption{(left): Observed velocity field of one IRDC from our sample. Contours are Herschel H$_2$ column density contours. (middle): Modelled velocity field obtained as described in the text. (right): Observed (blue solid) and modelled (red dashed) VSFs.}
         \label{spectra}
   \end{figure}

\section{Dense gas velocity fields and velocity structure functions}

We used the N$_2$H$^+$(1-0) line emission to derive the dense gas velocity field projected onto the plane of the sky. These velocity maps show a variety of emission properties, from linear velocity gradients to more complex velocity fields. To characterise these maps in a systematic and homogenous way, we have used velocity structure functions (VSF). The VSF we use is defined as $S_v(l)=median\left\{\left[v(x_i+l)-v(x_i)\right]^2\right\}$ where $x_i$ is the coordinates of a pixels, $l$ is the space lag between a pair of pixel. Note that the usual definition involves an average over all pairs of pixels as opposed to a median, but the median is less sensitive to the noise. Figure~2 (right) shows the square root of the VSF for  all clouds. It displays a rather large variety of morphologies, although one can visually define three groups of VSFs: i. linear; ii. plateaued; and  iii. oscillatory. The origin and connection between these different VSFs is not clear yet.  To get some more insights on the physical meaning of the observed VSFs, we modelled one specific filamentary IRDC showing an oscillatory VSF (see Fig.~3). We modelled its velocity field  following $v(z,r)=v_{sys}+r[\bigtriangledown v_r]+v_0\cos(2\pi z/\lambda)$ where $(r,z)$ are the radial and longitudinal pixel coordinates with respect to the main axis of the filament, $v_{sys}=46$\,km/s  is the systemic velocity of the cloud,  $\bigtriangledown v_r=1.1$\,km/s/pc is a constant velocity gradient in the radial direction, $v_0=0.5$\,km/s is the amplitude of the longitudinal oscillations, and $\lambda=2.0$\,pc is their wavelength. We then created a velocity map and constructed the modelled VSFs in both radial and longitudinal directions that we visually compared to the observed VSFs (see Fig.~3). The reasonable agreement suggests that the velocity field of that particular IRDC is dominated by a combination of radial collapse and pc-scale fragmentation. 

To make further progress in our understanding of VSFs, a similar analysis will be performed on all 27 clouds, and be compared to VSFs from cloud simulations.




\begin{thebibliography}{99}
\bibitem[2012] {C12} Crutcher, 2012, ARA\&A, 50,29
\bibitem[2007] {kt07} Krumholz \& Tan, 2007, ApJ,654, 304
\bibitem[2010] {m10} Molinari et al., 2010, A\&A, 518, 100
\bibitem[2009] {pf09} Peretto \& Fuller, 2009,A\&A, 505, 405
\bibitem[2013] {p13} Peretto et al., 2013, A\&A, 555, 112
\bibitem[2013] {s13} Smith et al., 2013,ApJ, 771, 24
\bibitem[1974]{ze74} Zuckerman \& Evans, 1974, 192, L149
\end{thebibliography}
\end{document}